# ELECTRONIC CONDUCTIVITY MEASUREMENTS IN SOLID ELECTROLYTES USING AN ION BLOCKING MICROELECTRODE: NOISE REJECTION BASED ON A MEDIAN FILTER

VEYIS GUNES[*], JEAN-YVES BOTQUELEN and ODILE BOHNKE

*Institut des Molécules et Matériaux du Mans (IMMM), Département des Oxydes et Fluorures*
*CNRS UMR 6283, Université du Maine, LUNAM, 72085 Le Mans Cedx 9, France*
*veyis.gunes@univ-lemans.fr*

A method of electronic conductivity measurement is presented. It combines two well known methods of electrochemistry: Cyclic voltammetry and Chronoamperometry. This DC technique uses the Hebb-Wagner approach to block ionic conduction (when steady state conditions are reached) and allows electronic conduction of solid electrolytes to be determined. In order to get short diffusion times, a micro contact is used as an ion blocking electrode. However, as the electronic conduction in electrolytes is and should be very low, the current is also very low, typically some tens of nanoamps. Thus, the heating system inevitably generates noise problems that are solved using a median filter. As opposed to other related work, our system allows the determination of the conductivities without any preliminary smoothing or fitting of the curves (since the noise is strongly reduced). Some results with oxygen ion conductors are also given.

## 1. Introduction

The determination of the partial electronic conduction in the total conduction (ionic + electronic) is essential to predict the quality of a material as an electrolyte (electrolytes should have mainly ionic conduction). In solid electrolytes which are $O^{2-}$ ion conductors, the electronic conductivity $\sigma_e$ varies with the oxygen activity of the gas in contact with the material (i.e., the oxygen partial pressure, as a first approximation). Such measurements are very important for the development of solid oxide fuel cells, oxygen sensors and gas separating membranes [Jang and Choi, 2002].

In the design of solid electrolytes for fuel cells, the oxygen activity limits of predominant ionic conduction have to be known. These limits determine the domain of electrolytic behavior. To determine this range, the electronic conductivity has to be measured and compared to ionic or total conductivity. In the design of oxygen sensors, the onset of electronic conduction limits the range of oxygen activity in which the sensor can operate, especially in low oxygen activities. The measurement of this conduction allows studies and characterization of these sensors [Sridhar and Blanchard, 1999].

Impedance spectroscopy is a technique widely used to determine the total conduction of materials, but it does not separate the ionic conduction from the electronic conduction. The determination of the electronic conductivity of a material as a function of oxygen activity ($a_{O2}$), can be done using different techniques. Lübke and Wiemhöfer [1999] have developed a modified method of the conventional Hebb-Wagner [Hebb, 1952, Wagner, 1956, Kim and Yoo, 2001, Shimonosono et al., 2004] method, which avoids the complex preparation of the partial pressures of oxygen in the gas stream. It is an electrochemical method which uses an ion blocking microelectrode. It allows doing measurements in a wide range of oxygen pressure ranging from $10^{-20}$ to $10^5$ atm, in a neutral atmosphere and continuously. These experiments are carried out in a wide temperature range depending on the materials investigated. In fuel cells, the temperature of the electrochemical cell can reach high temperatures such as 700 or 800°C. Since the current due to partial electronic conductivity are and should be very low (electrolytes should only have ionic conduction), the measurements are subjected to the noise generated by the heating system. In our laboratory, we have developed a system which eliminates this noise from the measurements by rejecting all and only the noisy measurements. This is done by adapting the cycle time of the temperature controller (controlled itself by a computer which also carries out data acquisition) to the sampling frequency and the application of a median filter to the measurements.

---

[*] Corresponding author.

As opposed to other related work (see for example Jang and Choi [2002], Kobayashi *et al.* [1997]), our measurements are reliable enough to avoid smoothing or fitting raw data as a preliminary to the calculation of the conductivities. The aim of this article is then to present an electronic conductivity measurement system which can be applied to materials used in solid oxide fuel cells and in oxygen sensors mainly. Then, we present the software we developed for the control of the system, data acquisition, signal processing and the conductivity determination. The problems we encountered in this type of acquisitions and the means used to solve them will also be presented. Finally, experimental results obtained on classic $O^{2-}$ ion conductors (YSZ) are presented.

## 2. The Hebb-Wagner Method

This method is based on an electrochemical cell (see Fig. 1) consisting in an ion-blocking electrode (platinum micro contact), an electrolyte (studied sample) and a reversible electrode (composed of CuO and $Cu_2O$ oxides, radius ≈ 5 mm). Practically, a reversible electrode is a non-polarizable electrode where charged species can easily go or come from [Lee, 2004]. The electrolyte is an $O^{2-}$ ionic conductor.

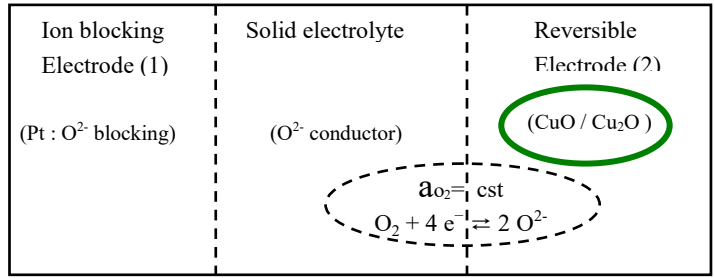

Fig. 1: Electrochemical cell, composed by an ion blocking electrode, a solid electrolyte and a reversible electrode, as suggested by Hebb-Wagner. At the second interface, the activity of $O_2$ is constant.

At the electrolyte/reversible electrode interface, the equilibrium is described by the Eq. (1):

$$2CuO \leftrightarrow Cu_2O + \frac{1}{2}O_2 \,. \tag{1}$$

The activity of oxygen is constant and the equilibrium implies an exchange of oxygen ions with electrons in order to keep oxygen activity constant at the interface (constant chemical potential µO2). When there is a flux of electronic charge carriers, inducing a current, a flux of oxygen ions occurs in the form of an equal and opposite current.

In order to get very short diffusion times Lübke and Wiemhöfer [1999] have proposed that the platinum electrode should be very narrow (we use a radius a≈100 µm). A schematic representation of such a cell is shown in Fig. 2. When a DC voltage is applied to the cell (which corresponds to a certain oxygen partial pressure on Pt contact), the O2- ions move to or from the blocking electrode (totally isolated from O2 from the ambient atmosphere with glue) depending on the sign of the voltage and the current through the sample is ionic and electronic. After a certain time, steady state conditions are reached and the remaining current becomes only electronic. The results (voltage, steady state current) are recorded and this is repeated for different voltages. Finally, electronic conductivities, as a function of oxygen partial pressure, are deduced.

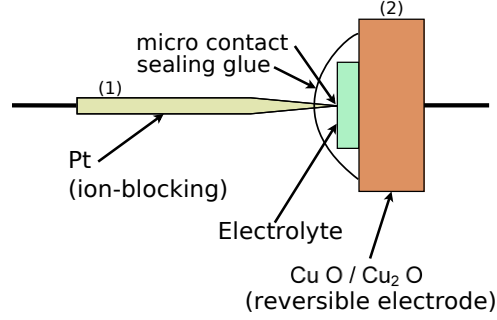

Fig. 2: Schematic representation of the cell. In our case, the radius of the reversible electrode is 5 mm. The micro contact has typically a radius of 100 μm.

The electronic conductivity $\sigma_e$ depends on the oxygen activity $a^1_{O_2}$ on the ion-blocking electrode. In order to vary this activity (i.e. the oxygen partial pressure), the following method is used (see Fig. 3). First, a DC voltage $E_{app}$ is applied to the cell (a few hundred millivolts is enough to generate oxygen activity values from $10^{-20}$ to $10^5$ atm. This creates a new value of $a_{O_2}$ on this electrode. Then there is diffusion of oxygen through the sample. After a delay, which depends on different factors, steady state conditions are reached. After that, the current and the associated voltage $E_{app}$ are recorded for the construction of the Hebb-Wagner curve (steady state current $I_{ss}$ as a function of the applied voltage $E_{app}$).

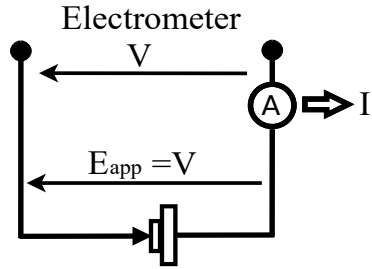

Fig. 3: Application of a DC voltage to the cell and measurement of the current by an ammeter (identified by A). When steady state conditions are reached, $I=I_{ss}$.

Usually, the Nernst equation, for the reaction $O_2 + 4\,e^- \rightleftharpoons 2\,O^{2-}$, is used in the following way:

$$E_{app} = E_1 - E_2 = \frac{RT}{zF} \ln\left(\frac{a^1_{O_2}}{a^2_{O_2}}\right). \qquad (2)$$

where $E_1$ and $E_2$ are, respectively, the galvanic potentials at electrodes 1 (platinum) and 2 (reversible electrode), $F$ is the Faraday's constant, $R$ is the universal gas constant, $T$ the temperature in Kelvin and $z$ is the number of involved electrons in the reaction (in our case: $z=4$).

Since the oxygen activity at the second interface ($a^2_{O_2}$) is constant, we can calculate the oxygen activity at the first interface ($a^1_{O_2}$) with the Nernst equation:

$$a^1_{O_2} = a^2_{O_2} \exp\left[\frac{4E_{app}F}{RT}\right]. \qquad (3)$$

Assuming that the surface of the micro contact has a hemispherical shape and that its radius, $a$, is much lower than the distance between it and the reversible electrode (i.e. the thickness of the electrolyte), which is true, one can calculate the electric resistance of the electrolyte [Rickert and Wiemhöfer, 1983], which is:

$$R = \frac{1}{2\pi a}\frac{1}{\sigma_e}. \tag{4}$$

This is a widely used (by semi-conductor technology) estimate [Mazur and Dickey, 1966] of resistance in a semi-infinite solid when a small-area contact is formed with a perfectly conducting hemispherical metal micro contact. In our case, the electrolyte represents the semi-infinite solid with an electronic conductivity $\sigma_e$. Thus, using the Hebb-Wagner curve (also called I-V curve), the electronic conductivity as a function of the applied voltage is given by:

$$\sigma_e(a_{O_2}^1) = \frac{1}{2\pi a}\frac{dI_{ss}}{dE_{app}}. \tag{5}$$

where $a$ is the radius of contact. In all the measurements presented in this article we have chosen $a=100$ μm).

Using the conventional method of Hebb-Wagner, Kobayashi et al.[1997] have used polynomial regressions to smooth this curve and then calculated the conductivities from these values. Jang and Choi [2002] fitted the raw data (current as a function of the applied voltage) with theoretical curves before calculating the partial electronic conductivities. Our method has the advantage that the measurements are good enough for calculating directly from our curves, except for a few points (near 0 V) for which $\Delta I_{ss} / \Delta E_{app}$ yields negative values for which logarithms are not defined. For these points (only one-contiguous) the $\Delta I_{ss} / \Delta E_{app}$ value is estimated to the mean value over the two neighboring points. These points are very rare and they occur only for the very low voltages.

## 3. A New Measurement Setup

In order to do all these measurements in an automated way, we have developed a workbench (see Fig. 4). The cell is placed in an oven and a voltage is applied on this cell by an electrometer (Keithley 6517A) which also carries out the measurements of current. This electrometer takes orders from the computer and sends the successive values of the current to the computer. We have developed a LabView application on this computer for these tasks. The user gives orders to the computer which gives orders to the temperature controller (TC) through a serial port (RS232). The TC allows the heating in the oven and transmits the temperature of the oven to the computer. The temperature rate and the duration of the dwell are controlled by our application through the TC. As the temperature in the cell is different from the oven temperature, another thermocouple is placed in the cell itself and its signal is transmitted to the computer through another data acquisition module connected to a second serial port.

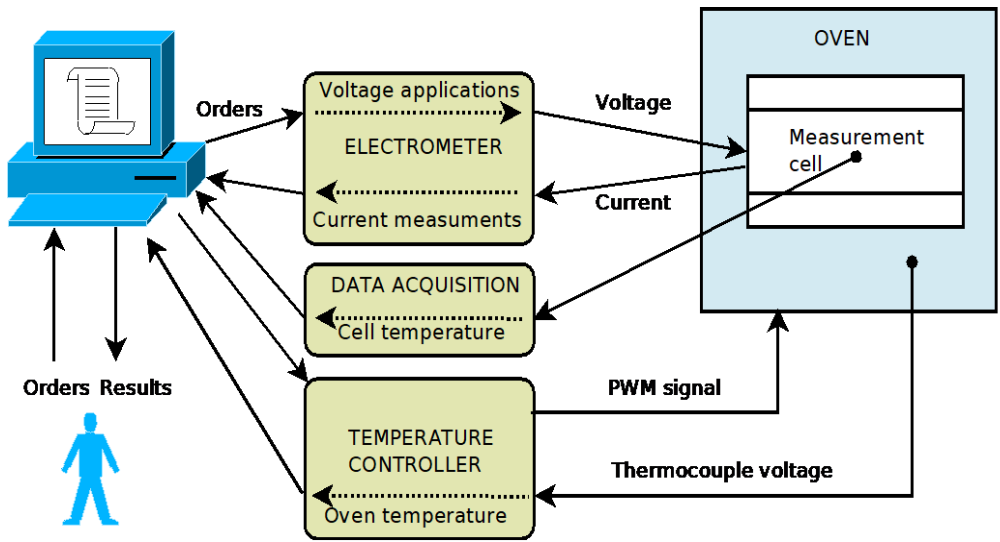

Fig. 4: Experimental setup for electronic conductivity measurements. Our LabView application controls the temperature of the cell, different applied voltages, data acquisition and data processing.

## 4. Software Development

The first task of the software (LabView application) is to give the set point temperature of the oven (calculated from the desired cell temperature) to the TC. When the oven temperature is stabilized, a second task is launched. It consists in stabilizing the desired cell temperature by adjusting the oven temperature through a PI (proportional-integral) controller.

Then, the measurement task itself begins. It consists in applying a DC voltage and measuring the current. It should be noted that the output impedance of the DC generator should be as low as possible and that the input impedance of the current entry should be as low as possible. We have chosen a Keithley 6517A which can execute these two tasks (see Fig. 3). After applying a DC voltage, the software reads, each second (default period: 1 s), the value of the current from the electrometer through a serial port. This is an over-sampling of the values given by the electrometer.

The software will detect the establishment of the steady state current, according to a threshold value $s$ (positive value, fixed by the user), over the variations. To be insensitive to weak variations (which can be due to noise), this threshold check is carried out only every $Np$ seconds and calculated from the mean values over $Nw$ seconds. These parameters are chosen by the user (we used: $Np=5$ $Nw=5$ and $s=3$ $nA$). The absolute difference of current between two successive checks is given by:

$$\Delta I = |I(t) - I(t-Np)| = \left| \frac{1}{Nw}\sum_{j=1}^{Nw} I(k+j) - \frac{1}{Nw}\sum_{j=1}^{Nw} I(k-Np+j) \right|. \qquad (6)$$

Thus, the current is considered to be in a steady state (see Fig. 5) when, for a check $k$, the following condition is satisfied:

$$\left| \frac{1}{Nw}\sum_{j=1}^{Nw} [I(k+j) - I(k-Np+j)] \right| \leq s. \qquad (7)$$

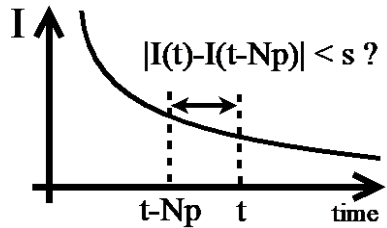

Fig. 5: Condition of steady state current. $Np$ is the delay between two successive threshold checks.

We usually used a threshold value of $s= 3.~10^{-9}$ A. When the steady state condition is reached, the current is calculated as the mean value of the current over $Nm$ measurements. Again, this parameter is free and it is chosen by the user, before launching the cycle of measurements. This steady state current value is a data of the I-V curve (or Hebb-Wagner curve). Finally, the software scans all voltages, with the voltage step given by the user, to get the complete I-V curve. During the measurements, the user has the possibility to change some measurement parameters, such as $Np$, $Nw$, $Nm$ and $s$. Also, all the current values (a file for each voltage applied) and all the I-V values are saved in files, "on the fly" (i.e. as they become available). At its launch, the software asks for the folder name, the extreme positive and negative voltages and the voltage step.

Two scanning methods are possible: down to negative extreme voltage, then up to positive extreme voltage and down to zero voltage (down-up-down procedure). In the same way, an up-down-up procedure is possible.

While the designed application is running, implicit parallel computing, inherent to LabView, permits the display of the current, a new value being added each second (since the sampling period of the electrometer is *200 ms*, this is an over-sampling). The algorithm for the measurement is as follows:

```
MEASUREMENT algorithm:
Inputs: SP_Oven: temperature set point for the oven
        T_Oven: real temperature in the oven
        SP_cell: temperature desired in the cell
        T_cell: real temperature in the cell
        Vstart, Vmin, Vmax, Vend, Vstep, Np, Nw, s, Nm
        Mfile: file name for the measurements
        Ki: integral coefficient of the PI controller
        Mode: Down-Up-Down (DUD) or UDU
        e: error allowed on temperatures (e=0.8°C)
Outputs: a set of measurements (Eapp, Iss) saved in Mfile
1:   Send (SP_Oven) to the Temperature controller (TC)
2:   Repeat
3:     Read T_Oven from the TC (using serial port 1)
4:     Wait (1 minute)
5:   Until T_Oven=SP_Oven (+/- e°C)
6:   Repeat
7:     Read T_cell from temperature module (using serial port 2)
8:     SP_Oven = SPOven + Ki(SP_cell-T_cell)
9:     If SP_Oven>1000 then SP_Oven=1000
10:    else if SP_Oven<0 then SP_Oven=0
11:    Send (SP_Oven) to the TC
12:    Wait (5 minutes)
13:    T_cell_old=T_cell
14:  Until (T_cell=SP_cell +/- e°C) and (|T_cell-T_cell_old|<0.15°C)
15:  If mode=DUD
16:    Decreasing voltages (i,Vstart,Vmin,Vstep,Np,Nw,s,Nm,Mfile)
17:    Increasing voltages (i+1,Vmin,Vmax,Vstep,Np,Nw,s,Nm,Mfile)
18:    Decreasing voltages (i+1,Vmax,Vend,Vstep,Np,Nw,s,Nm,Mfile)
19:  Else (i.e. mode=UDU)
20:    Increasing voltages (i,Vstart,Vmax,Vstep,Np,Nw,s,Nm,Mfile)
21:    Decreasing voltages (i+1,Vmax,Vmin,Vstep,Np,Nw,s,Nm,Mfile)
22:    Increasing voltages (i+1,Vmin,Vend,Vstep,Np,Nw,s,Nm,Mfile)
23:  End if
24:  Calculate conductivities from I-V curve (using Eq. 5)
25:  Draw conductivities as a function of oxygen activity (using Eq. 3)
```

The called algorithms are as follows:

```
INCREASING VOLTAGES algorithm:
Inputs:  ind: index of the voltage
         Umin, Umax, Ustep, Np, Nw, s, Nm
         Mfile: file name for the measurements
Outputs: a set of measurements (Eapp, Iss) saved in Mfile
1:   Repeat
2:     U=Umin + Ustep ind
3:     Apply voltage U (send U to Electrometer Keithley)
4:     Wait stabilization of I (s,Np,Nw) (uses Eq. 7)
5:     Iss = mean value of I (Nm: duration for integration)
6:     ind=ind+1
7:     Save (Eapp, Iss) in Mfile (add to Mfile)
8:   Until U ≥ Umax
```

```
DECREASING VOLTAGES algorithm:
Inputs:  ind: index of the voltage
         Umax, Umin, Ustep, Np, Nw, s, Nm
         Mfile: file name for the measurements
Outputs: a set of measurements (Eapp, Iss) saved in Mfile
1:   Repeat
2:       U=Umax - Ustep ind
3:       Apply voltage U (send U to Electrometer Keithley)
4:       Wait stabilization of I (s,Np,Nw) (uses Eq. 7)
5:       Iss = mean value of I (Nm: duration for integration)
6:       ind=ind+1
7:       Save (Eapp, Iss) in Mfile (add to Mfile)
8:   Until U ≤ Umin
```

### 5. Noise Rejection Using a Median Filter

An important drawback of the method is that the current signal is affected by a noise from the heating system (during the diffusion phase) which we identified in this study. The proposed solution rejects its disturbances on the measurements.

#### 5.1 Noise problem on the current curve

We carried out the measurements of electronic conductivity in a LLTO electrolyte (lithium lanthanum titanium oxide), at 637 °C, in air. As it can be seen below (see Fig. 6), a noise is present in the current signal and seems to be repeated every 10 seconds. This noise disturbs the steady state values of the current. In this example, the noisy value is about 100 nA higher than the value expected.

Since this noise disturbs also the I-V curve, conductivity values would also be affected. It was then of first importance to suppress this noise. However, it should be noted that only this noise should be suppressed. Thus, filtering or smoothing should be avoided (or at least, limited) in order to keep the useful information contained in the signals.

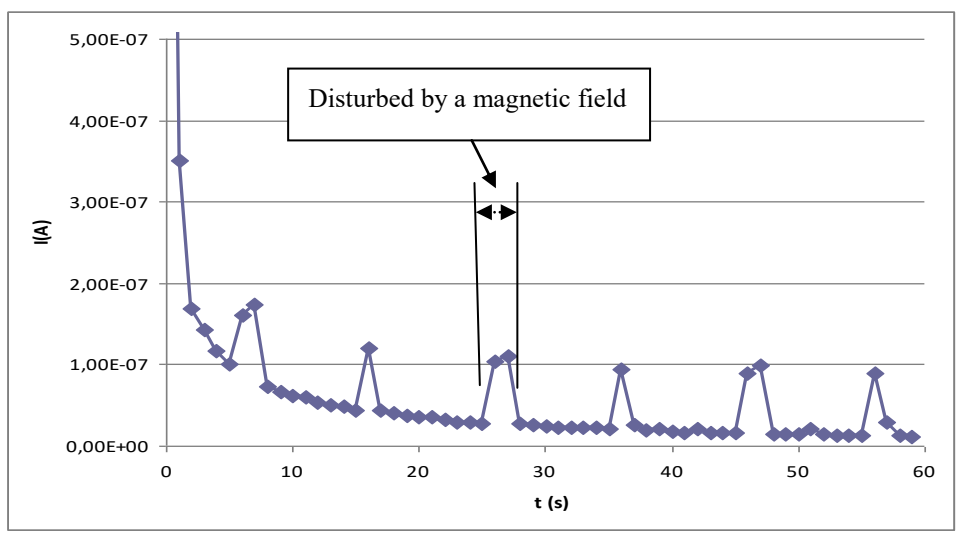

Fig. 6: Typical evolution of the current (as a function of time). The upper values of the current (every 10 seconds) are noisy measurements which should be eliminated.

## 5.2 Noise origin

Pulse-width modulation is a digital technique for varying the amount of power delivered to the heating element (resistor) of the oven. By varying the duration of the "on" state of the controller output, the amount of power delivered to the oven is adjusted and thus the temperature of the oven (and so of the cell too) is controlled.

However, after studying the origin of the noise, we discovered that the period of 10 seconds was due to the PWM (Pulse Width Modulation) functionality of the temperature controller (which had a duty cycle period of 10 seconds). The heating resistor (alas, wound) is the source of the noise. Each time the controller output is "on", the heating resistor of the oven is supplied by an AC voltage which creates a magnetic field in the cell. This field disturbs the measurements of current during diffusion, as well as when steady state conditions are reached. In the following, we explain how different solutions have been imagined and how we implemented a software solution.

## 5.3 Solutions

- Hardware solution discussion:

A cage of Faraday has been imagined, but practical and cheap materials (that can be shaped around the cell) can not tolerate high temperatures such as 900 °C or undergoes oxidation. A second solution would be to use forward/backward currents (in order to eliminate the resulting magnetic field) in doubled wired resistor elements. Unfortunately, the suppliers can not do this for a reasonable price. A third solution would be to insert an analog filter in the chain, but this solution shifts the value of current (always in the same way, since we noticed that it always adds or subtracts a contribution due to the magnetic field), so it is not an efficient solution. And finally, a DC power supply is not convenient as it will also create a constant magnetic field (depending on the set point value of the temperature).

- Software solution (by optimization of control parameters and the use of a median filter):

A "supply-off" measurement of the current is possible. This solution consists in setting the controller output "off" temporarily in order to check that the steady state conditions are reached and then to undertake a measurement of $I_{ss}$ (a mean value over $N_m$ measurements). This avoids creating a magnetic field in the cell. However, the temperature decrease, very fast at high temperatures such as 700°C, would disturb the measurements.

Another software solution, the technique of synchronization with the main supply, has been used but the magnitude of the noise was reduced by a ratio of 4 only. If this is associated with a median filter, the ratio increases to 8. The measurements are done while synchronized with the main supply. But unfortunately, this latter is not in the form of a pure sinusoidal voltage, thus a perfect synchronization is impossible. The solution is interesting but it can still be improved.

Finally, we solved this problem by improving and adapting device parameters (choice of a median type digital filter, part of our Keithley electrometer, a shorter cycle time for the TC and a suitable electrometer sampling period $T_s$). We used a sampling period $T_s$ in order to get the successive current values which fill the median filter of the electrometer. A higher sampling period also permits a higher accuracy measurement.

For a median filter with rank $R$, the median value is obtained by ordering the last $(2R+1)$ values and keeping the central value. This avoids taking into account, as less as possible, outliers, exceptional values, or in our case, values coming from disturbances. They remain exceptional as long as the corresponding duration of the heating (time during which TC output = "on": $Ton$) is lower than half the cycle time of the temperature controller $T_{tc}$. Thus, the median filter is efficient if the disturbance is shorter than half the time of the median filter window which is defined by $T_{WIN} = (2R+1) T_s$. This means that the disturbance should last less than ($T_{WIN}/2$), over a sliding window of $T_{WIN}$ seconds, meaning that:

$$T_{on} < \frac{(2R+1)T_s}{2}. \qquad (5.3)$$

We use a median filter in order to reject completely the disturbed values of the current. When it is done over a short period (for example, $T_{WIN}=2$ seconds), the median filtering does not change the mean value of the current (as the temperature is stable over these 2 seconds).

### 5.4 Efficiency of the solution

If the duration of the noise is less than half the temperature controller cycle time $T_{tc}$, (i.e. lower than $T_{on}$), the disturbed values (the noisy values) are completely rejected and replaced by precedent values. The median filter rejects all exceptional values by replacing them by a median value over the median filter window. So, this window must be as large as the controller cycle time $T_{tc}$, and the number of values read by the electrometer must correspond to a time equal to this cycle time. In other words, we have to choose $R$ and $T_s$ values in order to get $T_{WIN}$ as close as possible to $T_{tc}$.

In our case, with a rank of median filter $R=5$ and a sampling time of $T_s=200\ ms$, the width of the median filter window (see also Keithley documents) becomes: $T_{WIN} = (2R+1)\ T_s = 2.2\ s$. This is the nearest value to $T_{tc}$ (in our case, 2 s) that can be obtained. We checked the condition given by Eq. (5.3) for the different temperatures implied in our experiments. This condition is always respected (except during the temperature ramp during which we do not need to do current measurements).

### 5.5 Current curves of the proposed solution

In Fig. 7 (chronoamperometry), it has been shown that the magnetic field generated by the heating system no longer disturbs the current curve. As it can be seen in Fig. 8, the current curves obtained with our solution are far less noisy (considering that these currents are very low currents). It should be noticed that the end values of the current curves serve as an I-V point in the cyclic voltammetry curve. Of course, there may be still some noise on these curves, but it would no longer come from the identified source. If needed, other noise sources have to be considered. It should be noted that the shapes of these curves can vary, but this is tackled in the next sections.

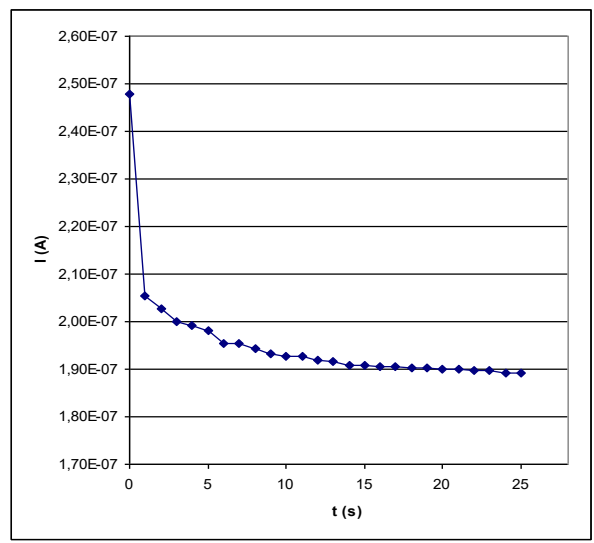

Fig. 7: Typical current shape obtained after noise treatment. The electrolyte is composed of $ZrO_2$ and 8 mol % of $Y_2O_3$ (Tosoh Zirconia) and is heated at a temperature of 700°C.

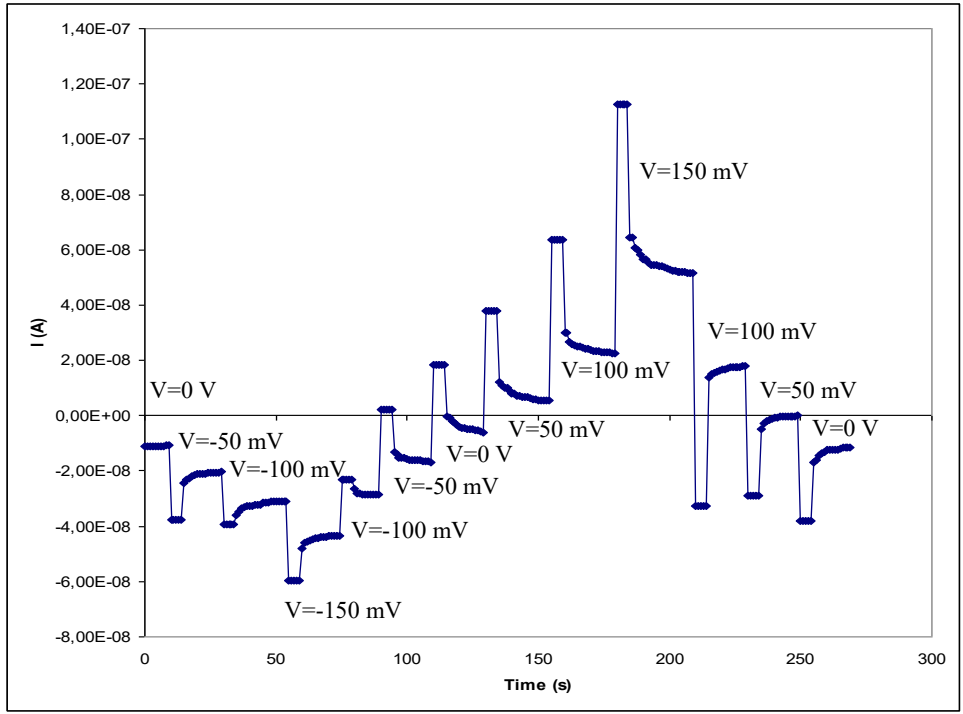

Fig. 8: Different current shapes obtained for different voltages at 650°C using a threshold of $10^{-9}$A. The electrolyte is a sample of Tosoh Zirconia.

## 6. Possible Improvements

Our system is, for now, more suited to relative measurements (qualitative measurements). It shows a good reproducibility which makes the system reliable. However, some improvements can be done. These improvements can lead to good quantitative measurements. Among these, we can mention: improving connections (shorter cables), improving the geometry of the micro contact (the error on *a* is about 10 %) and improving algorithms (for example in determination of conductivity).

## 7. Some Results on Yttrium Stabilized Zirconia (YSZ)

An I-V curve (cyclic voltammetry) is shown in Fig. 9. It represents the current as a function of the voltage applied to a typical and widely used electrolyte: the yttrium stabilized zirconium oxide (YSZ). We give the curves obtained for a mixture of $ZrO_2$ with 8 mol % of $Y_2O_3$ (Tosoh Zirconia). It should be noticed that the curve with the points is the curve used for the electronic conductivity determination. The curve often differs depending on decreasing or increasing voltages (we can notice this only after a cycle, for example a down-up-down cycle). This means that, for the low oxygen activity reached, a reduction has been initiated, i.e. oxygen atoms have been permanently removed. Experience showed that this difference is high when reduction has been far from the equilibrium. The same phenomenon can be observed at high oxygen activities but this is due to an oxidation which has been far from the equilibrium. Furthermore, the curve exhibits an S-like shape [Lee *et al.*, 2009] with an inflection point, implying a change of majority electronic charge carrier. As noticed by Jang and Choi.[2002], the curve becomes irreproducible above a certain temperature (depending on the material). This is probably due to the sample decomposition.

The final result of our system is the representation of the electronic conductivity as a function of oxygen activity (in log-log scale for practical reasons, see Fig. 10). These results give the value of $\sigma_e$ over a wide range of oxygen activity. Also, these curves allow (without smoothing or fitting with theoretical curves) the experimental determination of the slopes of $\sigma_{e^-}$ and $\sigma_p$ (the electronic conductivity by electrons and by holes, respectively) for electrolytes. They are calculated by linear regression over the values included in the corresponding range of activities. They correspond to the regions where the conduction is mainly due to electrons and holes (respectively in the low and high oxygen activity regions). These slopes of -1/6 and 1/6 (see Fig. 10) are close to the values given by the theory presented in the literature [Gao and Sammes, 2011, Reidy and Simkovich, 1993]. In the intermediate activity values (from $10^{-19}$ to $10^{-9}$), the measurements are too noisy (due to other sources than the heating system) to conclude.

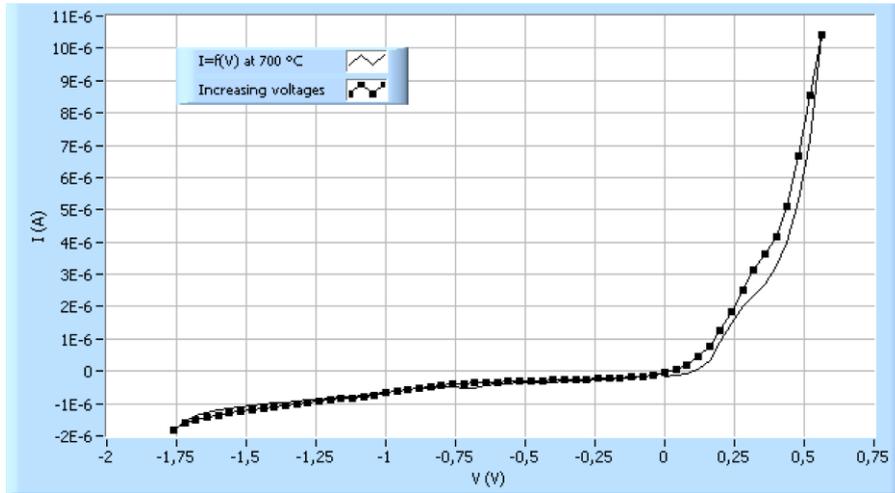

Fig. 9: Typical I-V curve at 700°C, obtained with the system described above. The sample used is a stabilized $ZrO_2$ electrolyte supplied by Tosoh (with 8 mol % of $Y_2O_3$).

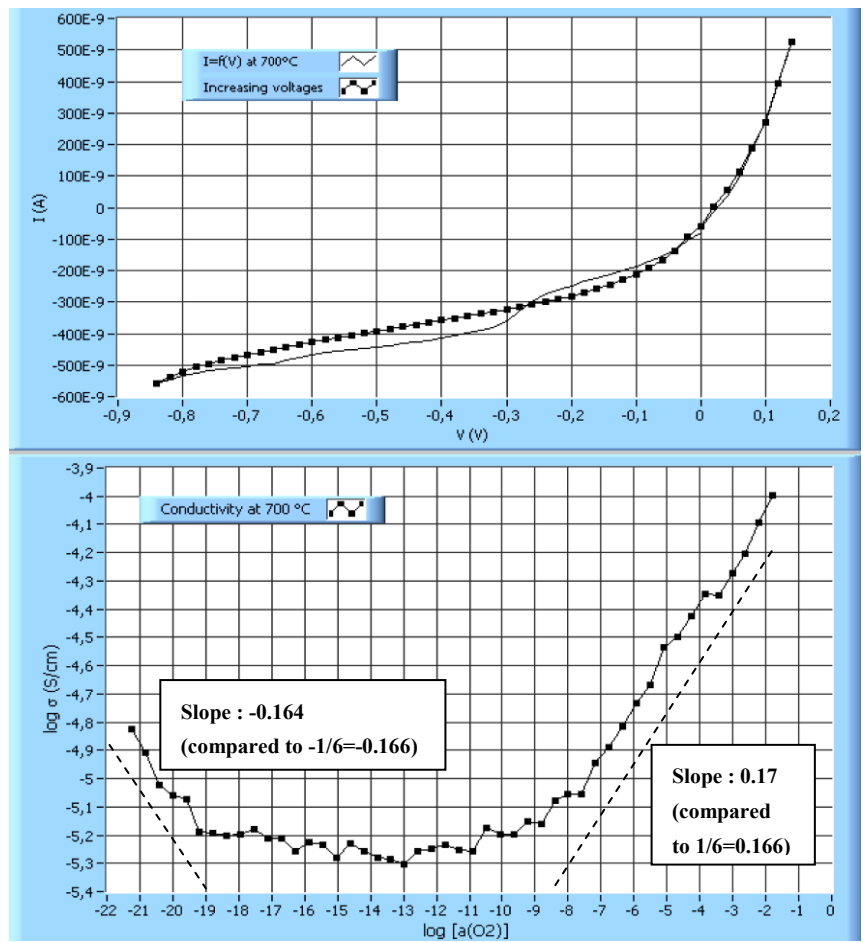

Fig. 10: Electronic conductivities calculated from the I-V curves of stabilized $ZrO_2$ (with 8 mol % of $Y_2O_3$). The slopes are consistent with the theoretical studies.

## 8. Attempt of Current Curves Analysis

The obtained curves can be summarized by Fig. 11. This latter represents a typical I-V curve, the accumulation of charges at low and high activities, the type of working of the involved devices (receptor or generator), and finally the shapes of the currents. These currents are full of information that can be analyzed, in order to study chemical and/or physical phenomena (e.g. the diffusion coefficient of $O^{2-}$ ions as a function of $O^{2-}$ oxygen ion concentration). We have drawn the most common types of curves over all measurements carried out, with different electrolytes. These latter are from YSZ, LLTO, $La_2Mo_2O_9$, and $BaCeO_3$ [Caldes *et al.*, 2012] families, mainly.

From these curves, it can be noticed that when a new voltage is applied, two stages can be seen; a first one during which the cell behaves mainly as a capacitor and thus undergoes a charge or a discharge and a second one in which it behaves mainly as an electrolyte and thus undergoes mainly a reduction or an oxidation (if the voltage is enough high). For these reasons, we have noted the flow of electrons, the charges of electrons, reduced oxygen ions and oxygen ion vacancies (with the Kröger-Vink notation). Further studies are needed. Some curves are more complex than those represented in the figure.

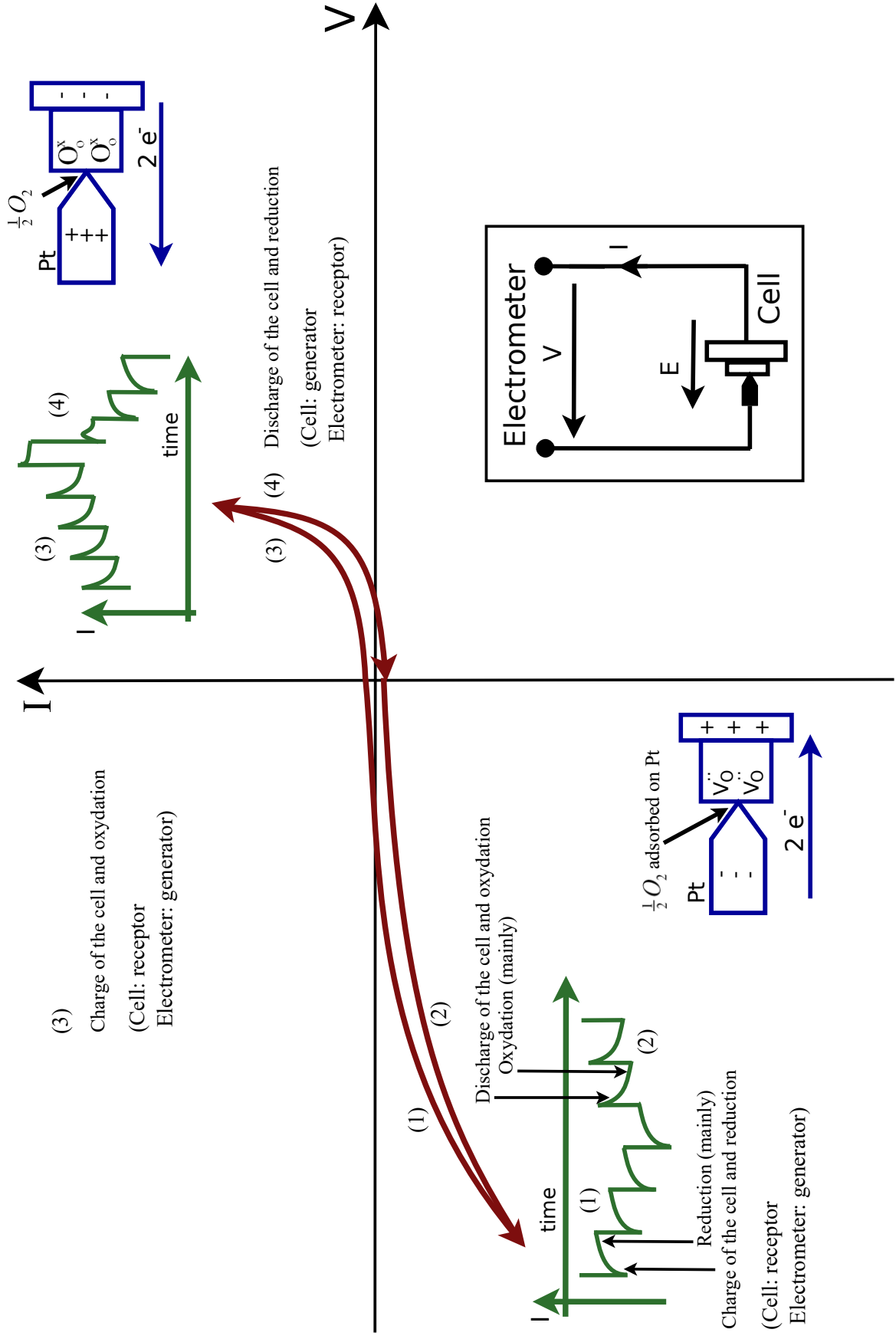

Figure 11: Current shapes during measurements (in nitrogen), and supposed distribution of electrical charges

# 9. Conclusions

In this work, a new electronic conductivity measurement system is proposed. It is based on the theory of Hebb-Wagner and the work of Lübke and Wiemhöfer. Our system is freed from the noise coming from the heating system by combining a median filter with adapted sampling parameters and temperature controller cycle time. The electronic conductivities are calculated from the I-V curves, with Yttrium stabilized $ZrO_2$ as an electrolyte example. As opposed to other related work, and since the remaining noise (due to other sources) is very low, these conductivities are calculated directly from the raw data, without preliminary smoothing or fitting with theoretical curves. Thus, our system constitutes a powerful tool to determine the electronic conductivity as a function of oxygen activity. These conductivities cover a wide range of oxygen activity with small voltages applied to a cell. An analysis of the shapes of the currents (for different voltages) is introduced. Improvements, including a better quantification of electronic conductivity and interpretation of the shapes of the obtained currents will be useful for further developments.